\documentclass[aps,prd,onecolumn,10pt,nofootinbib]{revtex4-2}

\usepackage{amsmath,amssymb}
\usepackage{graphicx}
\usepackage{xcolor}
\usepackage{placeins}
\usepackage{comment}
\usepackage{etoolbox}

\usepackage[colorlinks=true,linkcolor=blue,citecolor=blue]{hyperref}

\begin{document}

\setlength{\parskip}{6pt}
\setlength{\parindent}{0pt}
\linespread{1.05}

\title{Resonant Leptogenesis in a Two-Triplet Type-II Seesaw: A Dynamical Origin of Suppressed Lepton Flavor Violation}

\author{Avinanda Chaudhuri}
\email{avinandac@gmail.com}
\affiliation{Department of Physics,\\ Brahmananda Keshab Chandra College,\\ 111/2 B. T. Road, Kolkata-700108, India}

\date{\today}

\begin{abstract}

We investigate resonant leptogenesis in a two-triplet Type-II seesaw framework and demonstrate a coherent and predictive connection between neutrino mass generation, baryogenesis, and charge lepton flavor violation (LFV).

In the presence of quasi-degenerate scalar triplets, self-energy effects induce a resonant enhancement of the CP asymmetry, enabling successful baryogenesis at the TeV scale. We construct Yukawa couplings consistent with neutrino oscillation data and perform a comprehensive numerical analysis by solving the Boltzmann equations across a wide parameter space.

We find that viable solutions arise only within a restricted region characterized by near-resonant mass splittings and moderate-to-strong washout. In this regime, successful leptogenesis is achieved through resonant enhancement, which compensates for suppressed Yukawa couplings.

A key prediction of the framework is that the allowed parameter space dynamically favors small Yukawa couplings, leading to strongly suppressed LFV rates. The near-absence of observable LFV signals therefore emerges as a direct consequence of the dynamics responsible for baryogenesis.

Our results highlight a distinctive feature of the two-triplet Type-II scenario: the simultaneous realization of resonant enhancement and LFV suppression within a unified and testable framework.

\end{abstract}

\maketitle

\vspace{0.3cm}


\section{Introduction}

The origin of the observed baryon asymmetry of the Universe (BAU) remains one of the central open questions in particle physics and cosmology. Any successful explanation must satisfy the Sakharov conditions, namely baryon number violation, C and CP violation, and departure from thermal equilibrium~\cite{Sakharov1967}. Although these ingredients are present within the Standard Model (SM), they are insufficient to account for the observed asymmetry~\cite{Kuzmin1985, Fukugita1986, Buchmuller2005, Davidson2008, Drewes2017}, thereby motivating extensions beyond the SM.

Several mechanisms have been proposed to generate the BAU. Electroweak baryogenesis~\cite{Trodden1999, Morrissey2012} relies on a strongly first-order phase transition and additional sources of CP violation. Alternatively, the Affleck–Dine mechanism~\cite{Affleck1985, Dine1996} generates baryon asymmetry through the dynamics of scalar fields in the early Universe. Another particularly attractive possibility is baryogenesis via leptogenesis~\cite{Luty1992,  Covi1997, Plumacher1997}, where a lepton asymmetry is first generated and subsequently partially converted into baryon asymmetry through electroweak sphaleron processes~\cite{Arnold1987, Khlebnikov1988, Harvey1990}.

Leptogenesis is especially compelling due to its direct connection with neutrino mass generation. The smallness of neutrino masses can be naturally explained via the seesaw mechanism~\cite{Minkowski1977, GellMann1979, Yanagida1980, Mohapatra1980, Schechter1980}, which provides a framework for lepton number violation. In the canonical Type-I scenario, heavy right-handed neutrinos decay out of equilibrium, generating a CP asymmetry through the interference of tree-level and loop-level processes~\cite{Covi1997, Buchmuller2005BDP, Davidson2002}. 

Beyond the minimal Type-I framework, several variants have been extensively studied. Type-II leptogenesis~\cite{Ma1998, Antusch2004, AristizabalSierra2006, Hambye2003, Hambye2005} involves scalar triplets whose out-of-equilibrium decays generate a lepton asymmetry, with efficiency and CP violation strongly influenced by gauge interactions and flavor effects. Type-III leptogenesis~\cite{Hambye2004TypeIII, Strumia2009, Abada2008, Vatsyayan2022} utilizes fermionic triplets, where gauge-mediated interactions play a crucial role in determining the washout dynamics and overall efficiency.

A major challenge for standard thermal leptogenesis is the requirement of very high mass scales, typically above $10^9$ GeV~\cite{Davidson2002, Buchmuller2005}, which makes direct experimental tests difficult. This has motivated the exploration of low-scale realizations, including leptogenesis via neutrino oscillations in the $\nu$MSM framework~\cite{Akhmedov1998, Asaka2005, Canetti2013, Shuve2014}, as well as alternative scenarios based on extended or inverse seesaw mechanisms and resonant leptogenesis~\cite{Mohapatra1986, Kersten2007, Shaposhnikov2007, Drewes2015}.

In this context, resonant leptogenesis~\cite{ Pilaftsis2004, Pilaftsis2005, Dev2011} provides a particularly compelling mechanism. When two or more heavy states are nearly degenerate in mass, the CP asymmetry receives a resonant enhancement from self-energy contributions, allowing successful baryogenesis even at comparatively low scales. A consistent treatment of this regime requires careful handling of mixing and finite-width effects within non-equilibrium quantum field theory~\cite{Garny2010, Beneke2011, Kartavtsev2016}.

Motivated by these considerations, we investigate resonant leptogenesis within a two-triplet Type-II seesaw framework. While single-triplet realizations typically suffer from suppressed CP asymmetry, the presence of two quasi-degenerate scalar triplets allows for intrinsic CP-violating interference effects, naturally realizing resonant enhancement within the scalar sector. The same framework simultaneously generates neutrino masses and provides the conditions necessary for successful baryogenesis.

We analyze the interplay between scalar dynamics, CP asymmetry, and Boltzmann evolution, and explore the resulting phenomenological implications, including connections to charged lepton flavor violation (LFV). In this setup, the same Yukawa couplings govern neutrino masses, CP violation, and LFV observables, leading to non-trivial correlations among these sectors.

Our analysis reveals that successful leptogenesis is achieved only within a restricted region of parameter space characterized by near-degenerate triplet masses, $\Delta M / \Gamma \sim \mathcal{O}(1)$, and moderate-to-strong washout. In this regime, resonant enhancement compensates for suppressed Yukawa couplings, which in turn leads to strongly suppressed LFV rates. Consequently, the absence of observable LFV signals emerges as a natural consequence of the dynamics responsible for generating the BAU.

The paper is organized as follows. In Sec.~II, we present the scalar sector and neutrino mass generation. Sec.~III is devoted to the computation of CP asymmetry and its resonant enhancement. In Sec.~IV, we analyze the Boltzmann evolution and the interplay between production and washout effects. The numerical analysis and parameter space exploration are presented in Sec.~V, followed by a discussion of LFV constraints. We conclude in Sec.~VI.


\section{Scalar Sector and Neutrino Mass Generation}

In this section, we present the scalar sector of the model, consisting of two scalar triplet fields with hypercharge $Y=2$, as motivated in Type-II seesaw frameworks~\cite{Magg1980,Cheng1980,Lazarides1981,Mohapatra1981}.

The phenomenology of scalar triplets has been extensively studied in the literature, including their collider signatures, decay patterns, and implications for neutrino mass generation~\cite{Chun2003, Arhrib2011, Perez2008,Akeroyd2010,Chakrabarti1998}. Extensions with multiple scalar triplets further enrich the scalar sector and can lead to novel interference effects and modified decay channels~\cite{Chaudhuri2014,Grimus2008}. 

The presence of two scalar triplets in our framework allows CP violation through their interference, which is essential for generating the lepton asymmetry.

\vspace{0.2cm}

The triplet fields $\Delta_1$ and $\Delta_2$ can be written as $2\times2$ matrices:
\begin{equation}
\Delta_1 =
\begin{pmatrix}
\delta_1^+ & \sqrt{2}\,\delta_1^{++} \\
\sqrt{2}\,\delta_1^0 & -\delta_1^+
\end{pmatrix},
\qquad
\Delta_2 =
\begin{pmatrix}
\delta_2^+ & \sqrt{2}\,\delta_2^{++} \\
\sqrt{2}\,\delta_2^0 & -\delta_2^+
\end{pmatrix}.
\label{eq:triplets}
\end{equation}

After electroweak symmetry breaking, the neutral components of the triplets acquire vacuum expectation values (VEVs):
\begin{equation}
\langle \Delta_1 \rangle_0 =
\begin{pmatrix}
0 & 0 \\
v_1 & 0
\end{pmatrix},
\qquad
\langle \Delta_2 \rangle_0 =
\begin{pmatrix}
0 & 0 \\
v_2 & 0
\end{pmatrix}.
\label{eq:triplet_vev}
\end{equation}

The Standard Model Higgs doublet is given by:
\begin{equation}
H =
\begin{pmatrix}
H^+ \\
H^0
\end{pmatrix},
\end{equation}

with the vacuum expectation value:
\begin{equation}
\langle H \rangle_0 = \frac{1}{\sqrt{2}}
\begin{pmatrix}
0 \\
v
\end{pmatrix}.
\label{eq:higgs_vev}
\end{equation}

\subsection{Scalar Potential and Tadpole conditions}

The scalar potential involving the Higgs doublet $H$ and two scalar triplets $\Delta_1$ and $\Delta_2$ can be written as

\begin{align}
V(H,\Delta_1,\Delta_2) &= 
 \mu_H^2 \, H^\dagger H 
+ \lambda \, (H^\dagger H)^2 \nonumber \\[4pt]
&+ m_{\ell k} \, \mathrm{Tr}(\Delta_\ell^\dagger \Delta_k)
+ b_{\ell k} \, \mathrm{Tr}(\Delta_\ell^\dagger \Delta_k \Delta_\ell^\dagger \Delta_k) \nonumber \\[4pt]
&+ c_{\ell k} \, (H^\dagger H)\,\mathrm{Tr}(\Delta_\ell^\dagger \Delta_k)
+ d_{\ell k} \, \mathrm{Tr}(\Delta_\ell^\dagger \Delta_k)\,\mathrm{Tr}(\Delta_\ell^\dagger \Delta_k) \nonumber \\[4pt]
&+ e_{\ell k} \, H^\dagger \Delta_\ell \Delta_k^\dagger H \nonumber \\[4pt]
&+ f_1 \, \mathrm{Tr}(\Delta_1^\dagger \Delta_2)\,\mathrm{Tr}(\Delta_2^\dagger \Delta_1)
+ f_2 \, \mathrm{Tr}(\Delta_1^\dagger \Delta_1)\,\mathrm{Tr}(\Delta_2^\dagger \Delta_2) \nonumber \\[4pt]
&+ \left( \mu_k\, H^\dagger \Delta_\ell \tilde{H} + \text{h.c.} \right).
\label{eq:scalar_potential}
\end{align}

where $\tilde{H} = i \sigma_2 H^*$ is the charge-conjugated Higgs doublet and $\ell, k = 1,2$ runs over the two scalar triplets. Such scalar potentials involving triplet fields have been extensively studied in the literature~\cite{Chun2003,Arhrib2011}.


The scalar potential written above is not the most general. However, due to the smallness of the triplet vacuum expectation values (VEVs), several quartic terms are not numerically relevant for the scalar mass matrices. For simplicity, we neglect such terms in analytical expressions. Nevertheless, all numerical results presented in this work are obtained using the full scalar potential.

\vspace{0.2cm}

We assume the following orders of magnitude for the parameters in the scalar potential:
\begin{equation}
\mu_H^2 , \; m_{\ell k}\sim v^2, \quad 
\lambda, \; b_{\ell k}, \; c_{\ell k}, \; d_{\ell k}, \; e_{\ell k}, \; f_k \sim \mathcal{O}(1),
\end{equation}

\begin{equation}
|\mu_k| \ll v, 
\qquad 
\mu_k = |\mu_k| e^{i\phi_k}.
\end{equation}

The triplet VEVs satisfy:
\begin{equation}
|v_1|, |v_2| \ll v,
\end{equation}
in order to satisfy constraints from the $\rho$-parameter.

\vspace{0.2cm}

In our framework, we consider complex $\mu_1$ and $\mu_2$, which provide the necessary source of CP violation. These CP-violating effects from the triplet sector play a crucial role in generating the lepton asymmetry.


The tadpole equations of the scalar potential are obtained by minimizing the potential with respect to the scalar fields.

\vspace{0.2cm}

The Higgs tadpole condition is given by:
\begin{equation}
\mu_H^2 + \lambda v^2 
+ 2\sqrt{2}\, |\mu_1| v_1 \cos\phi_1 
+ 2\sqrt{2}\, |\mu_2| v_2 \cos\phi_2 = 0.
\label{eq:higgs_tadpole}
\end{equation}

The minimization of the scalar potential with respect to the triplet fields yields the following tadpole equations:

\begin{align}
m_{11} v_1 
+ \frac{1}{2}(c_{11} + e_{11}) v^2 v_1 
+ \frac{1}{2}(m_{12} + m_{21}) v_2 
\quad + \frac{1}{4}(c_{12} + c_{21} + e_{12} + e_{21}) v^2 v_2 
+ \sqrt{2}\, |\mu_1| v^2 \cos\phi_1 
= 0,
\label{eq:tadpole_v1}
\\[8pt]
m_{22} v_2 
+ \frac{1}{2}(c_{22} + e_{22}) v^2 v_2 
+ \frac{1}{2}(m_{12} + m_{21}) v_1 
\quad + \frac{1}{4}(c_{12} + c_{21} + e_{12} + e_{21}) v^2 v_1 
+ \sqrt{2}\, |\mu_2| v^2 \cos\phi_2 
= 0.
\label{eq:tadpole_v2}
\end{align}

The Higgs tadpole equation is recovered in the limit $v_\ell \to 0$ and $|\mu_\ell| \to 0$, which is consistent with our choice of parameter hierarchy.
\vspace{0.1cm}

After solving the Higgs tadpole condition, the parameters $\mu_1$ and $\mu_2$ can be determined from the triplet tadpole equations. In the limit $|\mu_\ell| \ll v$, triplet VEVs can be approximated as:
\begin{equation}
v_k \simeq \frac{\mu_k v^2}{M_{\Delta_k}^2},
\end{equation}
where $M_{\Delta_k}^2 \sim m_{kk}$ is triplet scalar mass.  This relation is characteristic of the Type-II seesaw mechanism and explains the smallness of neutrino masses~\cite{Magg1980,Cheng1980}.

\subsection{Charged Scalar Mass Matrix}

The mass matrix for the doubly charged scalar fields, in the basis $(\delta_1^{++}, \delta_2^{++})$, is given by:

\begin{equation}
M_{++}^2 =
\begin{pmatrix}
m_{11} + \dfrac{1}{2}(c_{11} + e_{11}) v^2 
& m_{12} + \dfrac{1}{2}(c_{12} + c_{21}) v^2 \\[8pt]
m_{12} + \dfrac{1}{2}(c_{12} + c_{21}) v^2 
& m_{22} + \dfrac{1}{2}(c_{22} + e_{22}) v^2
\end{pmatrix}.
\label{eq:doubly_charged_mass}
\end{equation}

\vspace{0.3cm}

The singly charged mass matrix in the basis $(h^+,\, \delta_1^+,\, \delta_2^+)$ is given by:

\begin{equation}
M^2_{+} =
\begin{pmatrix}
\mu_H^2 + \lambda v^2 
& \dfrac{v}{2\sqrt{2}}\left(4|\mu_1|\cos\phi_1 + e_{11} v_1 + e_{12} v_2 \right)
& \dfrac{v}{2\sqrt{2}}\left(4|\mu_2|\cos\phi_2 + e_{22} v_2 + e_{21} v_1 \right)
\\[10pt]

\dfrac{v}{2\sqrt{2}}\left(4|\mu_1|\cos\phi_1 + e_{11} v_1 + e_{12} v_2 \right)
& m_{11} + \dfrac{v^2}{4}(2c_{11} + e_{11})
& a_{12} + \dfrac{v^2}{4}(c_{12} + e_{21})
\\[10pt]

\dfrac{v}{2\sqrt{2}}\left(4|\mu_2|\cos\phi_2 + e_{22} v_2 + e_{21} v_1 \right)
& a_{12} + \dfrac{v^2}{4}(c_{12} + e_{21})
& m_{22} + \dfrac{v^2}{4}(2c_{22} + e_{22})
\end{pmatrix}.
\label{eq:neutral_mass_matrix}
\end{equation}

\vspace{0.2cm}

The above mass matrix has a zero eigenvalue corresponding to the charged Goldstone boson absorbed by the $W^\pm$ gage boson. This provides a consistency check of the scalar potential and the tadpole conditions. Both matrices are diagonalized by unitary transformations to obtain the physical mass eigenstates. Doubly charged scalars provide one of the most distinctive signatures of Type-II seesaw models, with characteristic decay modes into same-sign leptons, higgs and gauge bosons~\cite{delAguila2008}. In multi-triplet scenarios, mixing among triplet states can significantly modify these decay patterns.

\subsection{Yukawa Interactions and Neutrino Mass}

The $\Delta L = 2$ Yukawa interaction between the scalar triplets and the left-handed lepton doublets is given by:
\begin{equation}
\mathcal{L}_Y = \frac{1}{2} \sum_{k=1}^{2} 
Y^{(k)}_{ij} \, L_i^T \, C^{-1} \, i\sigma_2 \, \Delta_k \, L_j 
+ \text{h.c.},
\label{eq:yukawa}
\end{equation}

where $Y^{(k)}_{ij}$ are the symmetric Yukawa coupling matrices associated with the triplet $\Delta_k$. 
Here, $L_i$ denotes the left-handed lepton doublet and $i,j = 1,2,3$ are flavor indices.

After electroweak symmetry breaking, the neutrino mass matrix is generated from the above Yukawa interaction and is given by:
\begin{equation}
(M_\nu)_{ij} = \sum_{k=1}^{2} Y^{(k)}_{ij} \, v_k .
\label{eq:neutrino_mass}
\end{equation}

This structure is a generic feature of Type-II seesaw models, where neutrino masses arise directly from triplet Yukawa interactions~\cite{Hambye2012}.

For numerical analysis, we consider  the normal hierarchy of the neutrino mass spectrum. We set the lightest neutrino mass eigenvalue to zero. Using the experimentally observed central values of the neutrino mixing angles, the neutrino mass matrix can be reconstructed as:
\begin{equation}
M_\nu = U \, \hat{M}_\nu \, U^T,
\label{eq:pmns_reconstruction}
\end{equation}

The mixing matrix $U$ is identified with the Pontecorvo–Maki–Nakagawa–Sakata (PMNS) matrix~\cite{Pontecorvo1957, Maki1962}, whose parameters are determined from global analyses of neutrino oscillation data~\cite{NuFIT2024, JUNO2022, SNO2002, DayaBay2012}.
The diagonal neutrino mass matrix is $\hat{M}_\nu$ :
\begin{equation}
\hat{M}_\nu = \mathrm{diag}(0,\, m_2,\, m_3).
\end{equation}
where $m_2 = \sqrt{\Delta m_{21}^2}, \quad
m_3 = \sqrt{\Delta m_{31}^2}$.
The neutrino mass-squared differences are taken from global fits, with $\Delta m^2_{21}$ and $\Delta m^2_{31}$ determined with percent-level precision~\cite{NuFIT2024}.

and the PMNS matrix is 
\begin{equation}
U =
\begin{pmatrix}
c_{12} c_{13} & s_{12} c_{13} & s_{13} e^{-i\delta} \\
- s_{12} c_{23} - c_{12} s_{23} s_{13} e^{i\delta} 
& c_{12} c_{23} - s_{12} s_{23} s_{13} e^{i\delta} 
& s_{23} c_{13} \\
s_{12} s_{23} - c_{12} c_{23} s_{13} e^{i\delta} 
& -c_{12} s_{23} - s_{12} c_{23} s_{13} e^{i\delta} 
& c_{23} c_{13}
\end{pmatrix}.
\end{equation}

The values of the mixing angles $\theta_{12}$, $\theta_{13}$, and $\theta_{23}$ are taken from the latest global fits to neutrino oscillation data~\cite{NuFIT2024}. Recent measurements have significantly improved the precision of oscillation parameters, particularly $\theta_{12}$ and $\Delta m^2_{21}$. The above reconstruction allows us to relate the Yukawa couplings to low-energy neutrino data, providing a direct link between the scalar sector and neutrino phenomenology.


\section{CP Asymmetry}

The generation of the lepton asymmetry in this framework originates from the CP-violating decays of the scalar triplets $\Delta_i$ ($i=1,2$). These triplets decay via lepton-number-violating channels into lepton pairs and Higgs doublets,
\begin{align}
\Delta_i &\to \ell_\alpha \ell_\beta, \\
\Delta_i &\to H H,
\end{align}
where the latter is mediated by the trilinear scalar coupling $\mu_i$.

The total decay width of the triplets receives contributions from both channels and can be written as
\begin{equation}
\Gamma_i \simeq \frac{M_i}{8\pi} \, \text{Tr}(Y_i^\dagger Y_i) + \frac{|\mu_i|^2}{8\pi M_i}.
\end{equation}

The CP asymmetry is defined as
\begin{equation}
\epsilon_i = \frac{\Gamma(\Delta_i \to \ell\ell) - \Gamma(\Delta_i^\dagger \to \bar{\ell}\bar{\ell})}{\Gamma_i},
\end{equation}
and arises from the interference between tree-level and one-loop amplitudes~\cite{Pilaftsis1997}. Although the asymmetry is defined in the leptonic decay channel, its magnitude is controlled by both Yukawa couplings and scalar interactions, as reflected in the total decay width and loop contributions.

\vspace{0.3cm}
\noindent
\textbf{Origin of CP violation.} — In the two-triplet Type-II seesaw framework, CP violation arises from the interplay between Yukawa couplings and the scalar sector. The trilinear couplings are complex,
\begin{equation}
\mu_i = |\mu_i| e^{i\phi_i},
\end{equation}
and the physically relevant CP phase is the relative phase $\phi = \phi_2 - \phi_1$.

In addition, the Yukawa matrices $Y^{(1)}$ and $Y^{(2)}$ can contain complex phases. The CP asymmetry is governed by basis-invariant combinations involving both sectors, typically of the form
\begin{equation}
\mathrm{Im}\left[\mu_1 \mu_2^* \, \text{Tr}(Y_1 Y_2^\dagger)\right],
\end{equation}
which arise from interference between amplitudes mediated by different triplet states~\cite{Hambye2005, AristizabalSierra2006}.

\vspace{0.3cm}
\noindent
\textbf{Resonant enhancement.} — A distinctive feature of the present framework is the possibility of resonant enhancement when the two triplets are nearly degenerate in mass. In this regime, the dominant contribution to the CP asymmetry arises from self-energy diagrams, while vertex contributions are subleading and can be neglected.

The CP asymmetry can be expressed schematically as
\begin{equation}
\epsilon_i \propto 
\frac{\mathrm{Im}\left[\mu_1 \mu_2^* \, \text{Tr}(Y_1 Y_2^\dagger)\right]}
{\text{Tr}(Y_i^\dagger Y_i) + |\mu_i|^2/M_i^2}
\cdot
\frac{(M_i^2 - M_j^2) M_i \Gamma_j}
{(M_i^2 - M_j^2)^2 + M_i^2 \Gamma_j^2},
\end{equation}
which explicitly exhibits the Breit–Wigner structure characteristic of resonant leptogenesis~\cite{Pilaftsis2004, Dev2011}.

The resonance condition is achieved when
\begin{equation}
|M_1 - M_2| \sim \Gamma,
\end{equation}
leading to a maximal enhancement of the CP asymmetry. In this regime, the loop suppression is compensated by the near-degeneracy of the masses, allowing sizable asymmetry even for relatively small Yukawa couplings.

Away from resonance, the asymmetry exhibits the expected limiting behavior:
\begin{align}
\epsilon &\propto (\Delta M / \Gamma)^{-1}, \quad \text{for } \Delta M \gg \Gamma, \\
\epsilon &\to 0, \quad \text{for } \Delta M \ll \Gamma,
\end{align}
where $\Delta M = |M_1 - M_2|$.

\begin{figure}[t]
\centering
\includegraphics[width=0.48\textwidth]{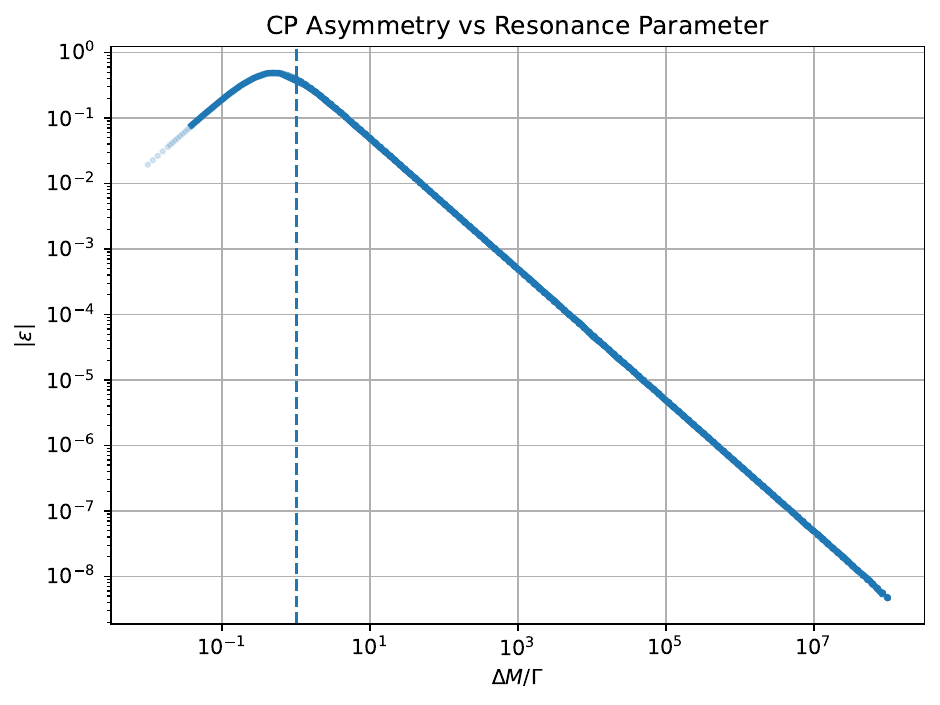}
\hfill
\includegraphics[width=0.48\textwidth]{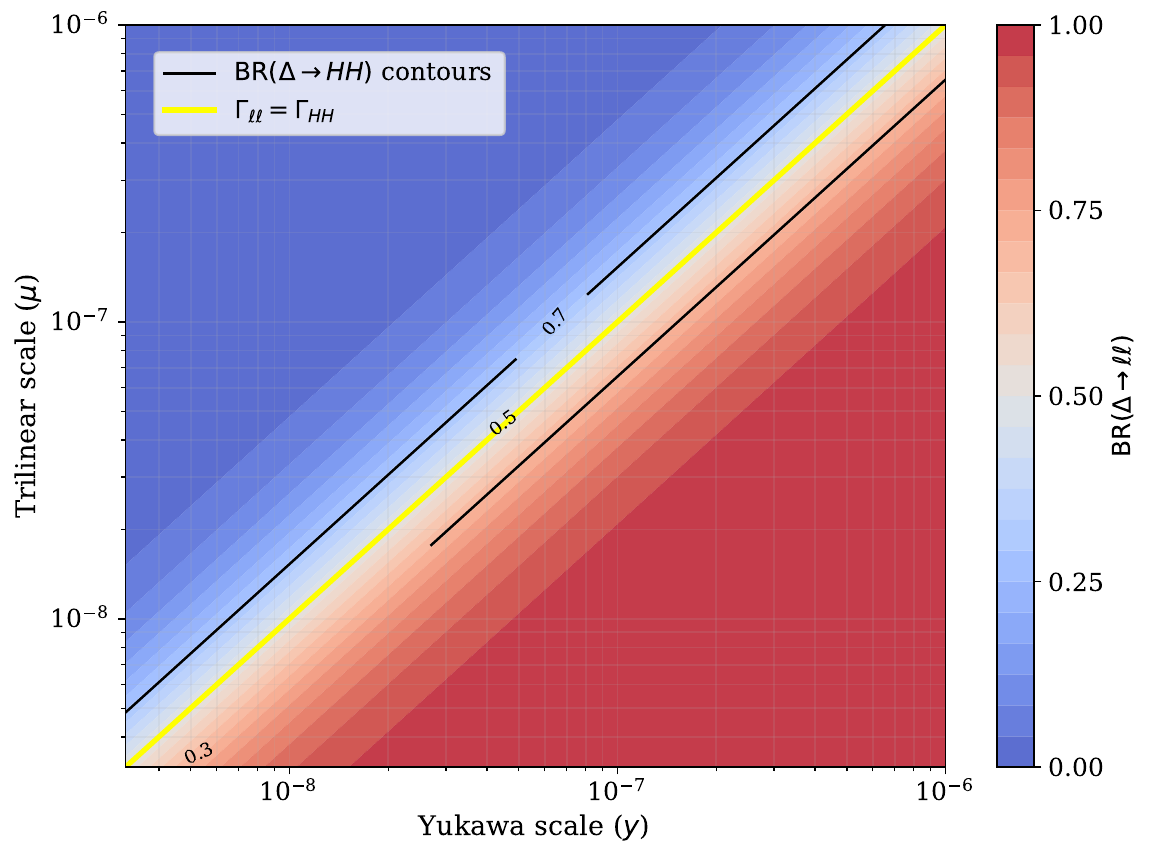}
\caption{
(a) CP asymmetry as a function of the resonance parameter $\Delta M / \Gamma$, showing a pronounced enhancement near $\Delta M \sim \Gamma$.
(b) Branching ratios of scalar triplet decays in the $(y,\mu)$ plane. The color map represents $\mathrm{BR}(\Delta \to \ell\ell)$, while contours correspond to $\mathrm{BR}(\Delta \to HH)$. The diagonal line indicates $\Gamma_{\ell\ell} = \Gamma_{HH}$.
The combined figure illustrates that successful leptogenesis requires both resonant enhancement and a balance between decay channels.
}
\label{fig:cp_and_br}
\end{figure}

This behavior is illustrated in Fig.~\ref{fig:cp_and_br}, where we show both the resonant enhancement of the CP asymmetry and the interplay between decay channels.

\vspace{0.3cm}

\begin{figure}[t]
\centering
\includegraphics[width=0.6\textwidth]{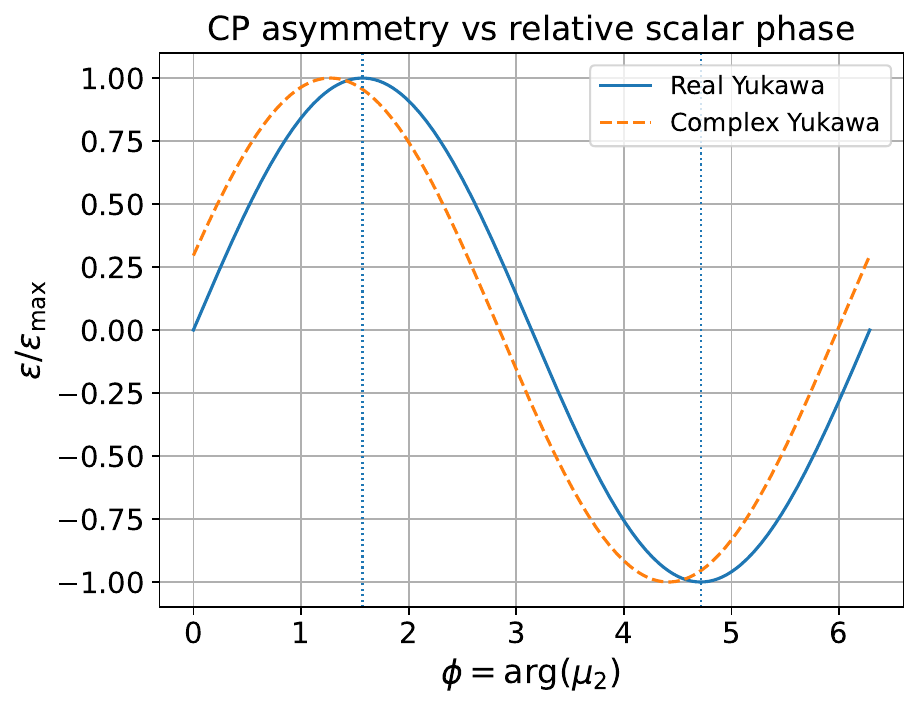}
\caption{
Dependence of the CP asymmetry on the relative phase $\phi = \arg(\mu_2)$.
The solid curve corresponds to real Yukawa couplings, exhibiting a purely sinusoidal behavior consistent with $\epsilon \propto \sin\phi$.
The dashed curve includes complex phases in the Yukawa sector, leading to a phase shift in the asymmetry.
Maximal CP violation occurs near $\phi \simeq \pi/2$.
}
\label{fig:phase_dependence}
\end{figure}

\noindent
\textbf{Phase dependence.} — The dependence of the CP asymmetry on the relative phase $\phi = \arg(\mu_2) - \arg(\mu_1)$ provides a direct probe of the CP-violating structure. For real Yukawa couplings, the asymmetry follows a sinusoidal behavior,
\begin{equation}
\epsilon \propto \sin\phi,
\end{equation}
indicating that CP violation is controlled by $\mathrm{Im}(\mu_1 \mu_2^*)$. This behavior is illustrated in Fig.~\ref{fig:phase_dependence}.

When complex phases are present in the Yukawa sector, the phase dependence is modified, leading to a shift in the position of maximal asymmetry, as shown in Fig.~\ref{fig:phase_dependence}.

\vspace{0.3cm}

\vspace{0.3cm}
\noindent
\textbf{Yukawa structure and parametrization.} — Unlike Type-I leptogenesis, where the Casas–Ibarra parametrization provides a general solution consistent with neutrino oscillation data~\cite{Casas2001}, the presence of multiple scalar triplets leads to a different structure of neutrino mass generation. 

In this work, we adopt a cancellation-based parametrization of the Yukawa matrices,
\begin{equation}
Y^{(1)} = \frac{M_\nu}{v_1} + \delta Y_1, 
\quad
Y^{(2)} = -\frac{M_\nu}{v_2} + \delta Y_2,
\end{equation}
where $\delta Y_{1,2}$ are small complex perturbations. This structure ensures that the leading contributions cancel, while controlled deviations reproduce the observed neutrino masses.

Such parametrizations are well motivated in multi-source neutrino mass models and allow sizable individual Yukawa couplings while maintaining consistency with low-energy data~\cite{Antusch2004, Hambye2012}.

\vspace{0.3cm}
\noindent
\textbf{Summary.} — The CP asymmetry in the two-triplet Type-II seesaw framework arises from the combined effect of scalar and Yukawa phases, and is resonantly enhanced in the quasi-degenerate regime. This provides a natural mechanism to generate the observed baryon asymmetry while maintaining compatibility with neutrino mass constraints.


\section{Boltzmann Evolution and Numerical Results}

Having established the origin and structure of the CP asymmetry, we now turn to its dynamical realization in the early Universe. The generated asymmetry is subject to production and washout processes governed by Boltzmann evolution. This provides the crucial link between microscopic CP violation and the observed baryon asymmetry.

\subsection{Boltzmann Equations}

We follow the evolution of the comoving number densities using the dimensionless variable $z = M_1/T$. The relevant quantities are the total triplet abundance $Y_\Delta$ and the $B-L$ asymmetry $Y_{B-L}$.

The Boltzmann equations are given by
\begin{align}
\frac{dY_\Delta}{dz} &= - \frac{z}{H(M_1)} \left[ \left( \frac{Y_\Delta}{Y_\Delta^{\rm eq}} - 1 \right) \gamma_D \right], \\
\frac{dY_{B-L}}{dz} &= - \frac{z}{H(M_1)} \left[ \epsilon \left( \frac{Y_\Delta}{Y_\Delta^{\rm eq}} - 1 \right) \gamma_D - W(z)\, Y_{B-L} \right],
\end{align}
where $\epsilon$ is the CP asymmetry derived in Sec.~III, $\gamma_D$ is the thermally averaged decay rate, and $W(z)$ encodes washout processes including inverse decays and $\Delta L = 2$ scatterings.

The thermally averaged decay rate is given by
\begin{equation}
\gamma_D = n_\Delta^{\rm eq} \frac{K_1(z)}{K_2(z)} \Gamma,
\end{equation}
where $K_{1,2}$ are modified Bessel functions and $n_\Delta^{\rm eq}$ is the equilibrium number density.

The Hubble expansion rate is
\begin{equation}
H(T) = \sqrt{\frac{4\pi^3 g_*}{45}} \frac{T^2}{M_{\rm Pl}},
\end{equation}
with $g_* = 106.75$.

The final baryon asymmetry is related to the $B-L$ asymmetry through sphaleron conversion~\cite{Kuzmin1985}:
\begin{equation}
Y_B = c_{\rm sph} \, Y_{B-L}, \qquad c_{\rm sph} \simeq -\frac{28}{79},
\end{equation}
and the baryon-to-photon ratio is given by $\eta_B \simeq 7.04\, Y_B$.

\subsection{Efficiency and Washout}

The efficiency of leptogenesis is controlled by the decay parameter
\begin{equation}
K = \frac{\Gamma}{H(T = M_1)},
\end{equation}
which quantifies the competition between decay processes and Hubble expansion.

Different regimes can be identified:
\begin{itemize}
\item $K \ll 1$: weak washout — inefficient production due to suppressed interactions,
\item $K \gg 1$: strong washout — asymmetry is generated but significantly erased,
\item $K \sim \mathcal{O}(1\text{--}10^4)$: optimal regime for successful leptogenesis.
\end{itemize}

In the present framework, the Yukawa couplings control both the CP asymmetry and the washout strength, while the trilinear coupling $\mu$ governs the total decay width. This leads to a non-trivial and highly constrained interplay between asymmetry generation and washout.

\subsection{Numerical Scan Strategy}

We perform a numerical scan over the parameter space defined by:
\begin{itemize}
\item the resonance parameter $\Delta M / \Gamma$,
\item the trilinear coupling scale $|\mu|$,
\item the Yukawa deformation scale,
\item CP-violating phases.
\end{itemize}

The parameter ranges and their physical roles are summarized in Table~I. For each parameter point, the Yukawa matrices are constructed to reproduce the observed neutrino mass matrix while allowing controlled complex perturbations that generate CP violation.

The Boltzmann equations are solved numerically to obtain the final baryon asymmetry $\eta_B$.

\subsection{Results and Physical Interpretation}

\begin{figure}[t]
\centering
\includegraphics[width=0.48\textwidth]{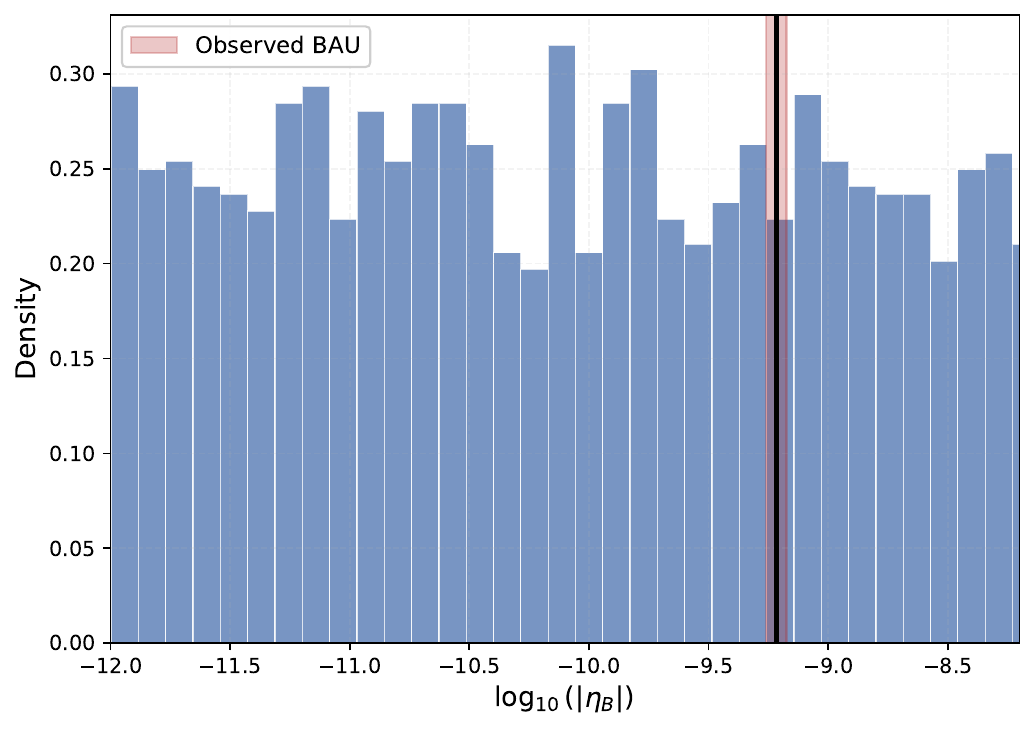}
\hfill
\includegraphics[width=0.48\textwidth]{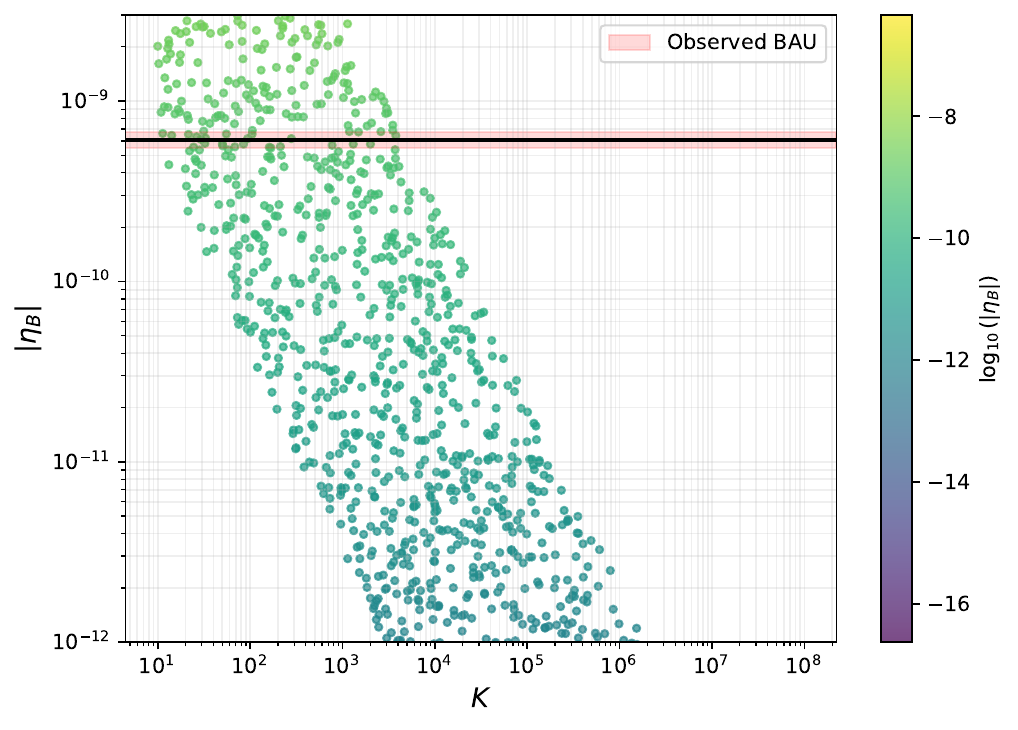}
\caption{
(a) Distribution of the generated baryon asymmetry $\eta_B$ from the parameter scan. The shaded band indicates the observed baryon asymmetry from Planck measurements, with the vertical line marking the central value.
(b) Dependence of the baryon asymmetry on the washout parameter $K$. The color map represents $\log_{10}(|\eta_B|)$, while the horizontal band denotes the observed baryon asymmetry.
Successful leptogenesis is realized in a restricted region of parameter space where both CP asymmetry and washout dynamics are appropriately balanced.
}
\label{fig:boltzmann_results}
\end{figure}

The results of the numerical scan are summarized in Fig.~\ref{fig:boltzmann_results}. A viable region consistent with the observed baryon asymmetry emerges as a narrow band in parameter space.

This structure reflects a non-trivial and highly constrained interplay between competing physical effects:
\begin{itemize}
\item For small $|\mu|$, the decay rate is suppressed, leading to inefficient generation of asymmetry.
\item For large $|\mu|$, the scalar decay channel dominates, reducing the branching ratio into leptons and suppressing the generated asymmetry.
\item Increasing the Yukawa couplings enhances CP violation but simultaneously strengthens washout processes.
\end{itemize}

As a result, successful leptogenesis is not generic, but arises only within a well-defined region where the competing effects are delicately balanced.

A key feature of the results is that successful baryogenesis occurs predominantly in the quasi-degenerate regime
\begin{equation}
\Delta M \sim \Gamma,
\end{equation}
which directly reflects the resonant enhancement discussed in Sec.~III.

\begin{figure}[t]
\centering
\includegraphics[width=0.8\textwidth]{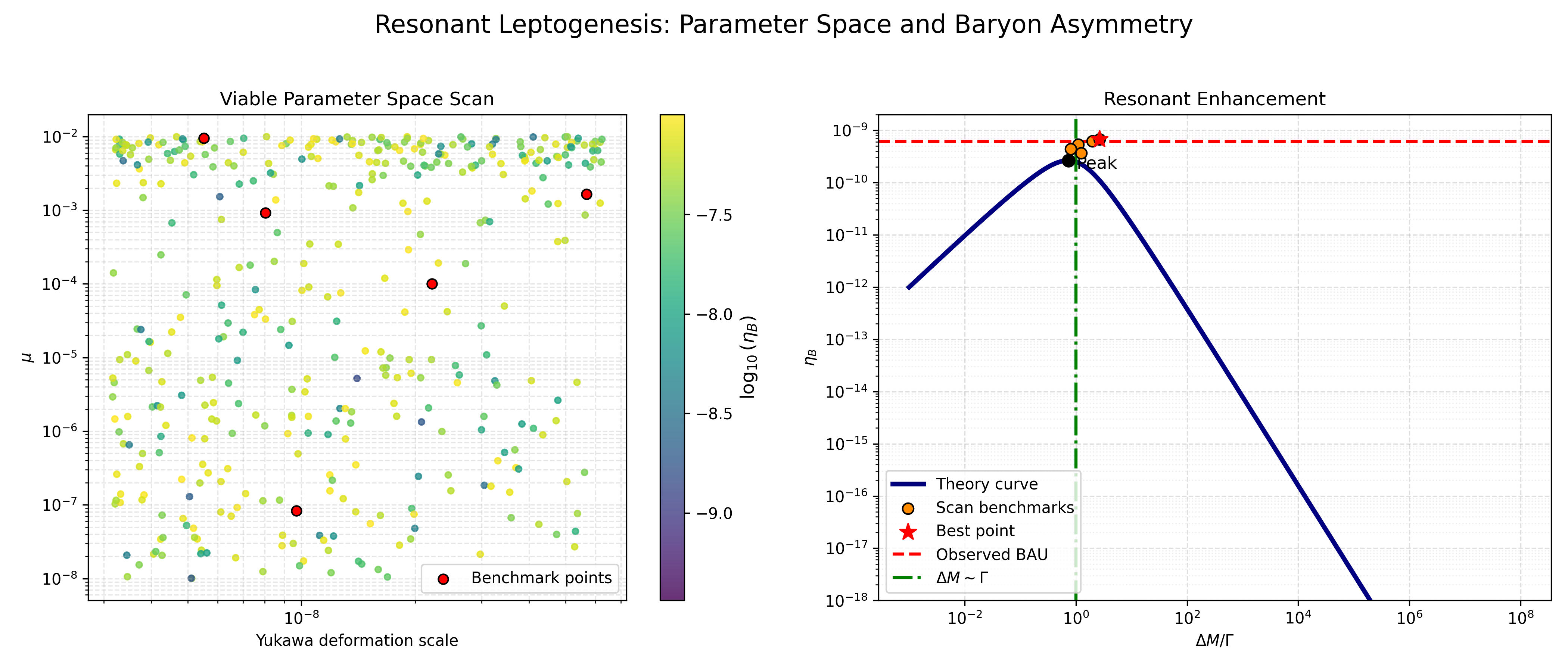}
\caption{
Left: Viable parameter space in the $(y,\mu)$ plane, colored by the generated baryon asymmetry $\eta_B$. Red points denote benchmark solutions consistent with the observed BAU.
Right: Baryon asymmetry $\eta_B$ as a function of the resonance parameter $\Delta M/\Gamma$. The blue curve shows the theoretical expectation, while orange points correspond to scan benchmarks. The red dashed line indicates the observed BAU, and the green dash-dotted line marks the resonance condition $\Delta M \sim \Gamma$.
}
\label{fig:parameter_scan}
\end{figure}

This behavior is illustrated in Fig.~\ref{fig:parameter_scan}, where we show the viable parameter space and the dependence of the baryon asymmetry on the resonance parameter.

The parameter ranges used in the numerical scan are summarized in Table~\ref{tab:parameters}, along with their physical roles in the framework. The scan is designed to capture the interplay between resonant enhancement, decay dynamics, and washout effects that govern successful leptogenesis.

In particular, the resonance parameter $\Delta M/\Gamma$ is varied around $\mathcal{O}(1)$ to probe the quasi-degenerate regime where the CP asymmetry is maximally enhanced, as discussed in Sec.~III. The trilinear couplings $\mu_{1,2}$ control the decay width of the scalar triplets and therefore play a crucial role in determining both the efficiency of asymmetry generation and the branching ratios into leptonic final states.

The Yukawa deformation parameters are introduced to generate CP violation while maintaining consistency with the observed neutrino mass matrix. Their magnitude directly affects both the size of the CP asymmetry and the strength of washout processes. The scan ranges are chosen to ensure compatibility with neutrino oscillation data while allowing sufficient freedom to explore viable leptogenesis solutions.

Overall, the parameter space is constrained by the simultaneous requirements of generating the observed baryon asymmetry, reproducing neutrino masses, and avoiding excessive washout, leading to a highly correlated and predictive structure.

\begin{table}[t]
\centering
\caption{Parameter ranges and their physical roles in the numerical scan.}
\begin{tabular}{lll}
\hline\hline
Parameter & Range & Role \\
\hline
$M_1$ & $10^3$--$5\times10^3$ GeV & Leptogenesis scale \\
$\Delta M/M_1$ & $10^{-8}$--$10^{-2}$ & Resonance control \\
$\Delta M/\Gamma$ & $\sim \mathcal{O}(1)$ & CP enhancement \\
$v_{1,2}$ & $10^{-8}$--$10^{-4}$ GeV & Neutrino mass \\
$|\mu_{1,2}|$ & $\ll v$ & Decay width \\
$|\delta Y|$ & $10^{-8}$--$10^{-2}$ & CP + LFV \\
Phases & $[0,2\pi]$ & CP violation \\
$\Gamma$ & computed & Out-of-equilibrium \\
$K$ & derived & Washout strength \\
\hline\hline
\end{tabular}
\label{tab:parameters}
\end{table}

The concentration of viable points near $\Delta M \sim \Gamma$ confirms that the observed baryon asymmetry in this framework is intrinsically tied to resonant leptogenesis. The resonant enhancement compensates for the relatively small Yukawa couplings required by neutrino mass constraints, allowing viable solutions at comparatively low scales.

\subsection{Benchmark Points}

To illustrate representative solutions, we select benchmark points from the viable region of parameter space. These are listed in Table~I.

\begin{table}[t]
\centering
\caption{Representative benchmark points consistent with the observed baryon asymmetry.}
\begin{tabular}{ccccc}
\hline\hline
Point & $\Delta M/\Gamma$ & $\eta_B$ & $\mu$ & Yukawa scale \\
\hline
A & 1.1085 & $5.32\times10^{-10}$ & $8.33\times10^{-8}$ & $9.71\times10^{-9}$ \\
B & 1.9963 & $6.19\times10^{-10}$ & $1.00\times10^{-4}$ & $2.21\times10^{-8}$ \\
C & 1.2369 & $3.64\times10^{-10}$ & $9.18\times10^{-4}$ & $8.04\times10^{-9}$ \\
D & 0.8034 & $4.43\times10^{-10}$ & $1.66\times10^{-3}$ & $5.69\times10^{-8}$ \\
E & 2.6675 & $6.74\times10^{-10}$ & $9.57\times10^{-3}$ & $5.52\times10^{-9}$ \\
\hline\hline
\end{tabular}
\label{tab:benchmarks}
\end{table}

The benchmark points listed in Table~\ref{tab:benchmarks} illustrate representative solutions within the viable parameter space that successfully reproduce the observed baryon asymmetry. A common feature of these points is that the resonance parameter $\Delta M/\Gamma$ lies close to $\mathcal{O}(1)$, confirming that efficient baryogenesis is realized in the quasi-degenerate regime where resonant enhancement of the CP asymmetry is operative. 

At the same time, the trilinear coupling scale $\mu$ and the Yukawa deformation parameters are adjusted such that the decay rate and washout strength remain in an optimal range. Smaller values of $\mu$ lead to suppressed decay rates and inefficient asymmetry production, while larger values enhance scalar decay channels and reduce the branching ratio into leptons. Similarly, increasing the Yukawa scale enhances CP violation but also strengthens washout effects. 

The benchmark solutions therefore demonstrate that successful leptogenesis arises from a delicate balance between resonant enhancement, decay dynamics, and washout processes. This highlights the predictive nature of the two-triplet Type-II framework, where the requirement of generating the observed baryon asymmetry imposes correlated constraints on both the scalar and Yukawa sectors. 

Overall, the numerical results demonstrate that the observed baryon asymmetry can be successfully generated within a tightly constrained region of parameter space, where resonant enhancement, controlled Yukawa structure, and moderate washout act in a coordinated manner. This reinforces the predictive nature of the two-triplet framework.The same Yukawa structure that governs leptogenesis also induces charged lepton flavor violation, providing an important phenomenological probe of the framework.


\section{Lepton Flavor Violation}

In the Type-II seesaw framework, lepton flavor violation (LFV) arises from the same Yukawa interactions responsible for neutrino mass generation. In the present setup with two scalar triplets $\Delta_{1,2}$, the relevant interactions are given by
\begin{equation}
\mathcal{L}_Y = Y^{(a)}_{\alpha\beta} \, L_\alpha^T C i\sigma_2 \Delta_a L_\beta + \text{h.c.}, \quad a=1,2,
\end{equation}
where $Y^{(a)}$ are symmetric Yukawa matrices.

After electroweak symmetry breaking, the neutrino mass matrix is generated as
\begin{equation}
M_\nu = Y^{(1)} v_{\Delta_1} + Y^{(2)} v_{\Delta_2}.
\end{equation}

In our framework, the Yukawa matrices are constructed via a cancellation-based parametrization, allowing the observed neutrino masses to emerge from a partial cancellation between the two triplet contributions. While this structure permits sizable individual Yukawa couplings in principle, the requirement of successful resonant leptogenesis imposes strong constraints on their effective magnitude.

\subsection{LFV Observables}

The Yukawa interactions induce LFV processes such as $\mu \to e\gamma$, $\mu \to 3e$, and $\mu$–$e$ conversion in nuclei. Among these, the radiative decay $\mu \to e\gamma$ provides the most stringent constraint. The current experimental upper bound from the MEG II experiment is
\begin{equation}
\text{BR}(\mu \to e\gamma) < 6 \times 10^{-14}.
\end{equation}

In this framework, the branching ratio receives contributions from both triplets and scales as
\begin{equation}
\text{BR}(\mu \to e\gamma) \propto 
\left| \sum_{a=1,2} \frac{(Y^{(a)\dagger} Y^{(a)})_{e\mu}}{M_a^2} \right|^2.
\end{equation}

Importantly, LFV observables depend on quadratic combinations of the Yukawa matrices and therefore do not inherit the cancellation structure present in the neutrino mass matrix. As a result, LFV provides an independent probe of the Yukawa sector.

\subsection{Correlation with Leptogenesis}

A key feature of the present framework is the tight connection between LFV and leptogenesis. As shown in Sec.~III and Sec.~IV, successful baryogenesis via resonant leptogenesis requires suppressed decay widths,
\begin{equation}
\Gamma_a \propto \text{Tr}(Y^{(a)} Y^{(a)\dagger}) M_a,
\end{equation}
ensuring out-of-equilibrium decays and avoiding strong washout.

This condition dynamically favors small Yukawa couplings across the viable parameter space. Since LFV branching ratios scale as quartic powers of the Yukawa couplings,
\begin{equation}
\text{BR} \sim \frac{|Y|^4}{M^4},
\end{equation}
this leads to a strong suppression of LFV observables.

This suppression is not imposed by hand, but emerges as a direct consequence of the leptogenesis requirement, establishing a non-trivial link between high-scale baryogenesis and low-energy flavor physics.

\subsection{Numerical Results}

\begin{figure*}[!t]
\centering

\includegraphics[width=0.48\textwidth]{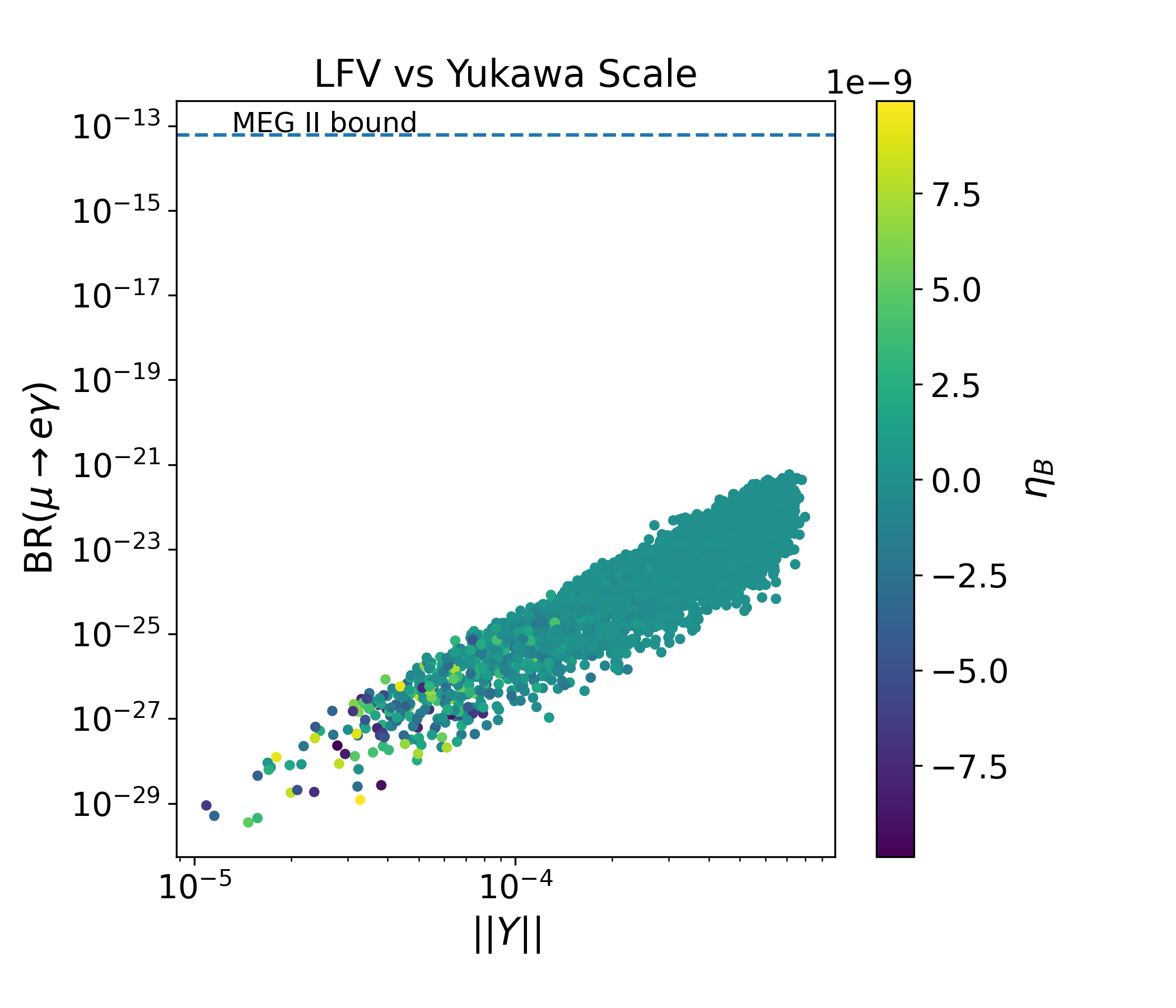}
\hfill
\includegraphics[width=0.48\textwidth]{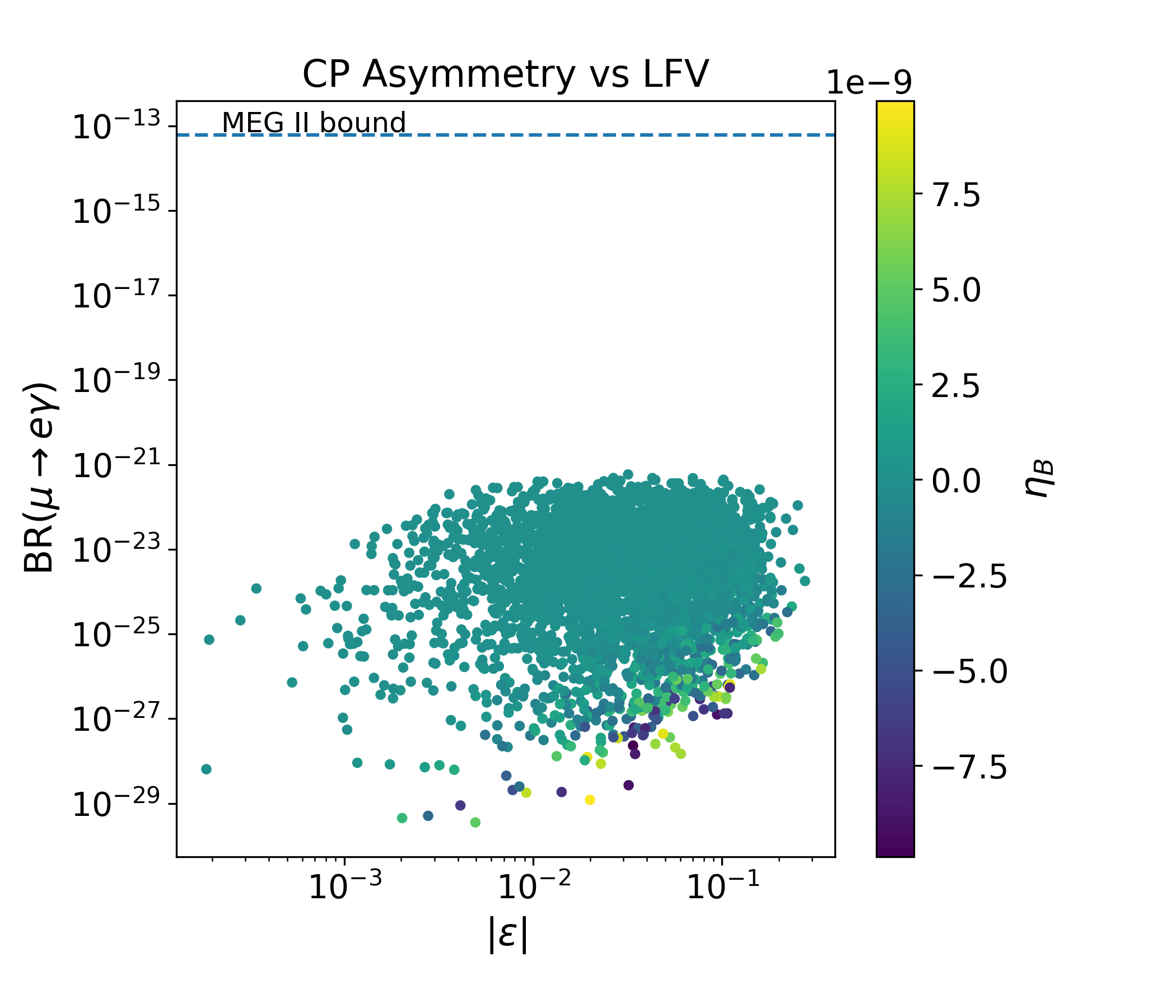}

\vspace{0.3cm}

\includegraphics[width=0.48\textwidth]{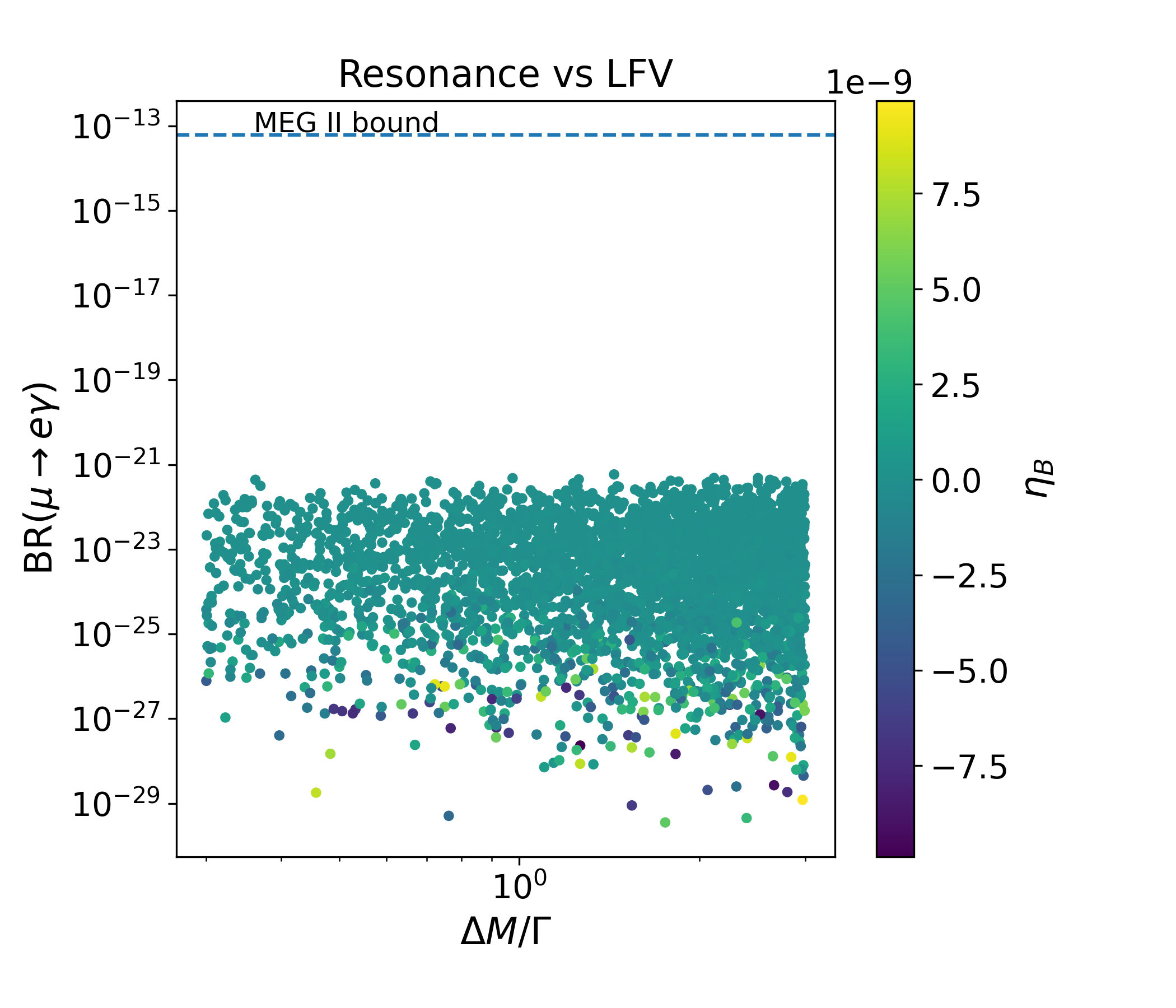}
\hfill
\includegraphics[width=0.48\textwidth]{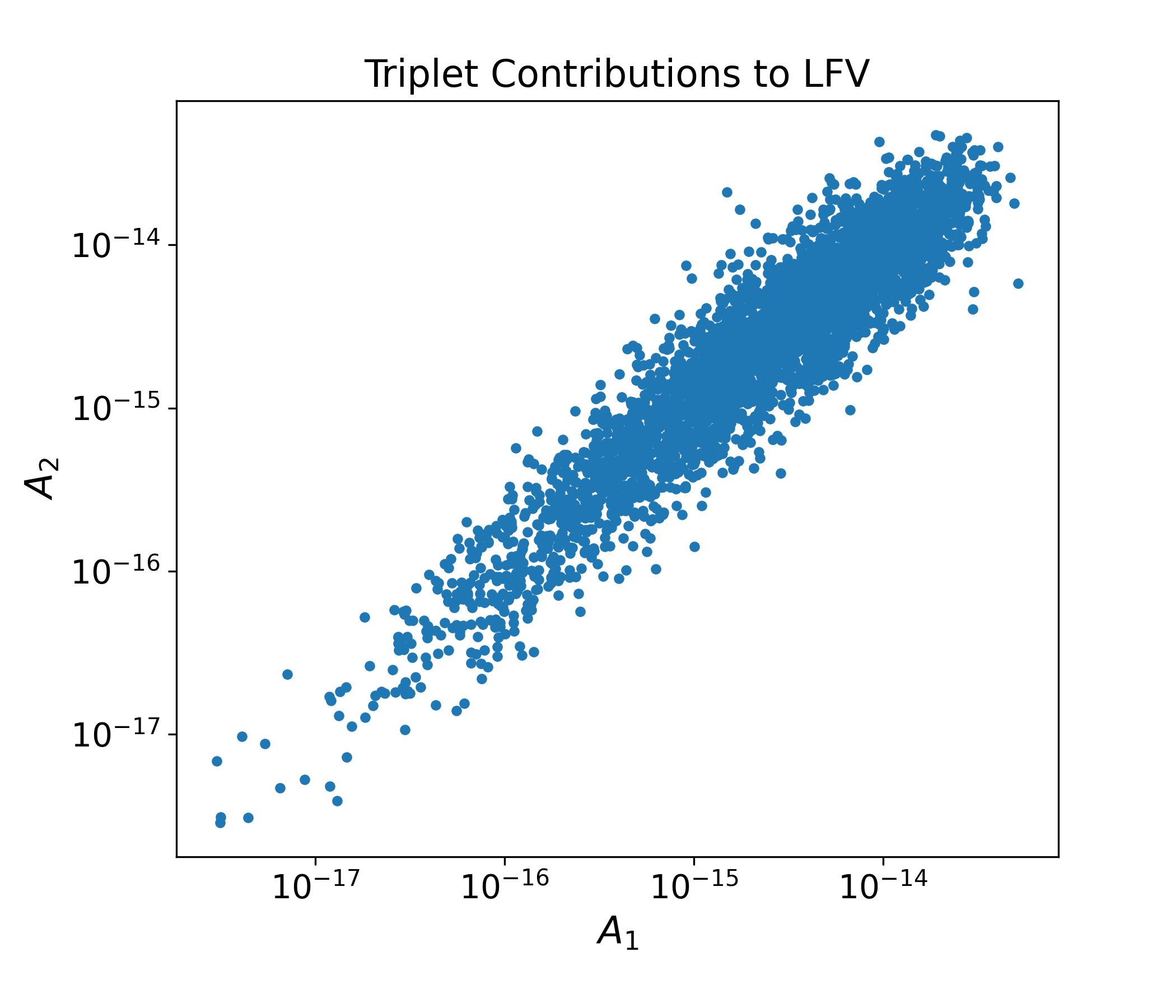}

\caption{
\textbf{Comprehensive view of lepton flavor violation in the two-triplet Type-II seesaw framework.}
(a) Branching ratio of $\mu \to e\gamma$ as a function of the effective Yukawa norm, showing the scaling $\text{BR} \propto ||Y||^4$, with all viable points lying well below the current experimental bound.
(b) LFV as a function of the CP asymmetry parameter, demonstrating that successful leptogenesis does not require enhanced LFV.
(c) Dependence of LFV on the resonance parameter $\Delta M/\Gamma$, indicating that resonant enhancement is largely decoupled from LFV observables.
(d) Correlation between the contributions of the two triplets to LFV amplitudes, showing the absence of fine-tuned cancellations.
}
\label{fig:lfv}
\end{figure*}

The interplay between LFV and leptogenesis is illustrated in Fig.~\ref{fig:lfv}.

Fig.~\ref{fig:lfv}(a) shows a clear correlation between LFV rates and the Yukawa scale, with $\text{BR} \propto ||Y||^4$. All points consistent with the observed baryon asymmetry lie well below the current experimental bound~\cite{MEG2016, MEGII2024}, typically in the range
\begin{equation}
\text{BR}(\mu \to e\gamma) \sim 10^{-29} - 10^{-22}.
\end{equation}

Fig.~\ref{fig:lfv}(b) demonstrates that large CP asymmetries do not require enhanced LFV, confirming that successful leptogenesis can be achieved without increasing Yukawa couplings.

Fig.~\ref{fig:lfv}(c) shows that LFV observables exhibit little dependence on the resonance parameter $\Delta M/\Gamma$, indicating that resonant enhancement of the CP asymmetry is largely decoupled from LFV.

Finally, Fig.~\ref{fig:lfv}(d) illustrates the contributions of the two triplets to LFV amplitudes. The strong correlation observed indicates that LFV does not rely on fine-tuned cancelations between different contributions.

\subsection{Implications}

The results reveal a distinctive and robust prediction of the framework: LFV observables are strongly suppressed throughout the parameter space, consistent with successful leptogenesis.

This suppression arises dynamically from the requirement of out-of-equilibrium decays and moderate washout, which enforce small Yukawa couplings. Consequently, the model predicts that LFV signals are likely to remain below the sensitivity of current experiments, despite the presence of new scalar states at relatively low scales.

This provides a clear phenomenological signature: the absence of observable LFV, in contrast to many other beyond-the-Standard-Model scenarios, is not accidental but a direct consequence of the leptogenesis dynamics.


\section{Discussion}

In this work, we have investigated resonant leptogenesis within a two-triplet Type-II seesaw framework, focusing on the interplay between neutrino mass generation, CP violation, baryogenesis, and charged lepton flavor violation (LFV). A central outcome of our analysis is the emergence of a highly non-trivial and predictive correlation among these sectors, governed by a common set of Yukawa interactions and scalar dynamics. This distinguishes the present framework from conventional resonant leptogenesis scenarios, where LFV and CP asymmetry are typically correlated through large Yukawa couplings.

The presence of two scalar triplets enables a cancelation-based parametrization of the neutrino mass matrix, in which the observed masses arise from a partial cancelation between the contributions of the two triplets. While this structure allows sizable individual Yukawa couplings in principle, our numerical analysis shows that the requirement of successful leptogenesis dynamically selects a restricted region of parameter space characterized by suppressed effective Yukawa norms.

The generation of the baryon asymmetry is driven by the resonant enhancement of the CP asymmetry in the quasi-degenerate regime $\Delta M \sim \Gamma$, where self-energy contributions dominate. In this regime, the Breit–Wigner enhancement compensates for the loop suppression, allowing sizable CP asymmetry even in the presence of small Yukawa couplings. The Boltzmann analysis further demonstrates that successful baryogenesis is realized only within a narrow region of parameter space, where the CP asymmetry and washout effects are optimally balanced.

A key implication of this dynamics is its impact on low-energy flavor observables. Since the same Yukawa couplings control both CP asymmetry and washout processes, the requirement of out-of-equilibrium decays naturally favors small Yukawa values. As LFV branching ratios scale as quartic powers of the Yukawa couplings, this leads to a strong suppression of LFV rates across the viable parameter space. Importantly, this suppression does not rely on fine-tuned cancelations but emerges as a direct consequence of the leptogenesis requirement.

Another notable feature of the framework is the decoupling between resonant enhancement and LFV observables. Although the CP asymmetry is highly sensitive to the resonance condition $\Delta M/\Gamma$, LFV rates exhibit little dependence on this parameter. This indicates that the mechanism responsible for enhancing the CP violation operates largely independently of low-energy flavor violation.

The numerical results further reveal that viable solutions are not uniformly distributed, but instead cluster within a tightly constrained region of parameter space where the competing effects of CP asymmetry and washout are balanced. This highlights the predictive nature of the framework: once the observed baryon asymmetry is imposed, both the Yukawa structure and the scalar sector parameters become strongly constrained.

From a phenomenological perspective, the model leads to a distinctive prediction: the absence of observable LFV signals in current experiments, despite the presence of new scalar states at relatively low scales. This stands in contrast to many extensions of the Standard Model where sizable LFV rates are expected. Future improvements in experimental sensitivity may probe the upper edge of the predicted LFV range, but the bulk of the parameter space remains well below current bounds.

Overall, the two-triplet Type-II seesaw framework provides a minimal and predictive realization of resonant leptogenesis in the scalar sector, in which neutrino mass generation, CP violation, and LFV are intrinsically linked. The resulting correlations offer a coherent and testable picture, connecting early Universe dynamics to low-energy observables.

\vspace{1cm}

\noindent{\textbf{Acknowledgement}}

The author expresses deep gratitude to her parents for continuous support throughout the work and to Professor Biswarup Mukhopadhyaya for introducing her to this fascinating field of research.


\end{document}